# Real time observation of a stationary magneton

Emmanouil Markoulakis*, Antonios Konstantaras, John Chatzakis, Rajan Iyer, Emmanuel Antonidakis

*Hellenic Mediterranean University former Technological Educational Institute of Crete, Department of Electronics Engineering, Computer Technology Informatics & Electronic Devices Laboratory, Romanou 3, Chania, 73133, Greece*
*Corresponding author: markoul@chania.teicrete.gr*

The magnetic dipole field geometry of subatomic elementary particles like the electron differs from the classical macroscopic field imprint of a bar magnet. It resembles more like an eight figure or else joint double quantum-dots instead of the classical, spherical more uniform field of a bar magnet. This actual subatomic quantum magnetic field of an electron at rest, is called Quantum Magnet or else a Magneton. It is today verified experimentally by quantum magnetic field imaging methods and sensors like SQUID scanning magnetic microscopy of ferromangets and also seen in Bose-Einstein Condensates (BEC) quantum ferrrofluids experiments. Normally, a macroscale bar magnet should behave like a relative giant Quantum Magnet with identical magnetic dipole field imprint since all of its individual magnetons collectively inside the material, dipole moments are uniformly aligned forming the total net field of the magnet. However due to Quantum Decoherence (QDE) phenomenon at the macroscale and macroscopic magnetic field imaging sensors limitations which cannot pickup these rapid quantum magnetization fluctuations, this field is masked and not visible at the macroscale. By using the relative inexpensive submicron resolution Ferrolens quantum magnetic optical superparamagnetic thin film sensor for field real time imaging and method we have researched in our previous publications, we can actually make this net magneton field visible on macroscale magnets. We call this net total field herein, Quantum Field of Magnet (QFM) differentiating it therefore from the field of the single subatomic magneton thus quantum magnet. Additionally, the unique potential of the Ferrolens device to display also the magnetic flux lines of this macroscopically projected giant Magenton gives us the opportunity for the first time to study the individual magnetic flux lines geometrical pattern that of a single subatomic magneton. We describe this particular magnetic flux of the magneton observed, quantum magnetic flux. Therefore an astonishing novel observation has been made that the Quantum Magnetic Field of the Magnet-Magneton (QFM) consists of a dipole vortex shaped magnetic flux geometrical pattern responsible for creating the classical macroscopic N-S field of magnetism as a tension field between the two polar quantum flux vortices North and South poles. A physical interpretation of this quantum decoherence mechanism observed is analyzed and presented and conclusions made showing physical evidence of the quantum origin irrotational and therefore conservative property of magnetism and also demonstrating that ultimately magnetism at the quantum level is an energy dipole vortex phenomenon. ©2019 The Authors. Published by Elsevier B.V. This is an open access article under the CC BY-NC-ND license (http://creativecommons.org/licenses/by-nc-nd/4.0/).

## 1. INTRODUCTION

A novel magnetic field physical sensor, quantum magnetic optic device is used to demonstrate and validate for the first time a correlation between the quantum net vortex field existing on macroscopic ferromagnets described in our previous work [1] and the familiar classical N-S macroscopic magnetic field imprint obtained, showing that macroscopic magnetism to be a Quantum Decoherence (QDE) effect. Therefore providing us a potentially important link between the quantum and macrocosm and possible enhancing our understanding over our physical world and magnetism in general. Similar quantum field images were previously reported independently in Bose-Einstein condensate quantum ferrofluids [21] *(i.e. ferrofluid close to absolute zero temperature)* and SQUID magnetic microscopy of ferromagnets [25], but never a valid correlation and observation was made with macroscopic magnetism and describing an existing quantum decoherence mechanism responsible for the transition from the quantum net magnetic field, thus quantun magnet to the macroscopic classical. This thin film magnetic optic lens can detect and display in real time this otherwise unobservable quantum field with conventional macroscopic sensors. A condensed soft quantum matter thin film, magnetic field sensor like the ferrolens, gives us a holographic imprint of the actual causality quantum magnet field of permanent magnets that more condensed macroscopic solid matter sensors are not able to show due to quantum decoherence phenomena limitations. This is not due an intrinsic property of the thin film ferrolens as we have established in our previous two publications referenced in this article but exclusively externally induced by the magnetic field under observation. This net effect quantum vortex field we experimentally discovered in every macro dipole magnet, is potentially the cause and responsible for the creation of its macroscopic classical axial field imprint we are all familiar with shown for example by macro field sensors like the iron filings experiment. The same or similar to the field vortices observed in Bose-Einstein condensate quantum ferrofluids.

The fact that this same quantum vortex field can be observed by a lens (i.e. Ferrolens) also at the macroscopic level and shown existing as a net result in every macro ferromagnet or electromagnet is remarkable and potentially important. The ferrolens acts like a quantum microscope and our method gives us potentially a unique opportunity for the first time to map the magnetic flux geometry and dynamics of the single elementary Magneton-Quantum Magnet thus the stationary magnetic field of a single electron. The whole magnet acts like a giant macroscopic stationary magneton when observed through the Ferrolens.

## 2. METHOD

A Ferrolens[1] quantum imaging device [1][2][13], (https://tinyurl.com/y2cgp59x), was used and its similar to aqueous magnetic film, magnetophotonic properties also independently reported by others [3,4]. This superparamagnetic [5–7] thin film optical lens exhibits minimal magnetic quantum decoherence [8–12] and has a sub-micron spatial resolution of the magnetic field under observation. The magnetic object under investigation is placed above or under the ferrolens at the center either with its pole facing the lens, polar-field (https://tinyurl.com/yctntnjc) view or at its side, side-field (https://tinyurl.com/y26yuru5) view although other configurations are also possible. The ferrolens can be activated using artificial lighting either by a light emitting diode (LED) ring around the perimeter of the lens or a single light source. In the case of the LED lighting, a wire-frame holographic pattern of the quantum field of the magnet is shown whereas if single light source is used, a cloud pattern of the field is displayed. Many types of materials of permanent magnets of different shapes were examined, a small sample we are presenting herein, ranging from neodymium Nd magnets, ferrite magnets as well electromagnets, all showing the same QFM vortex field geometry as shown below in fig.1.

---

[1] https://tinyurl.com/y4np83fn

## 3. RESULTS AND DISCUSSION

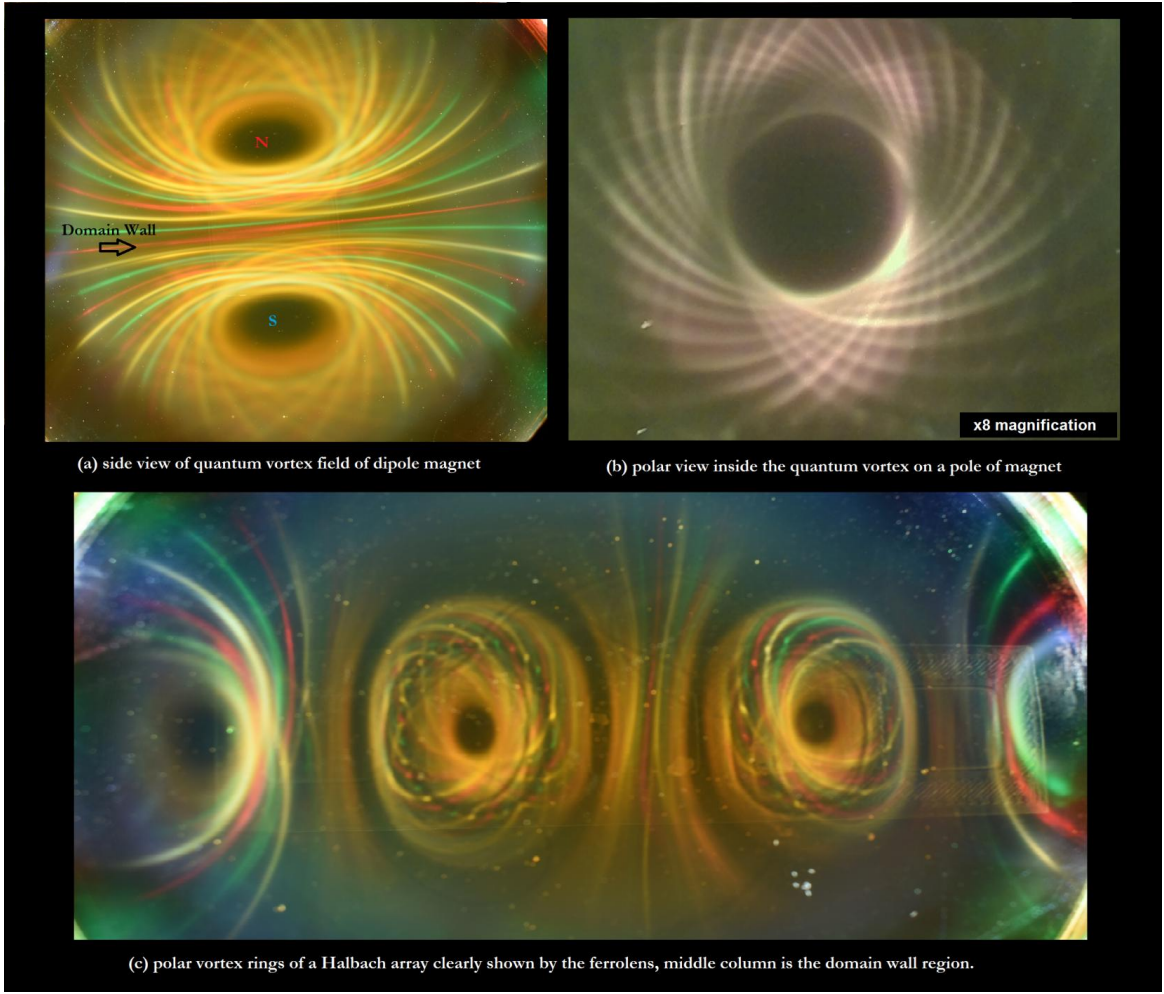

(a) side view of quantum vortex field of dipole magnet

(b) polar view inside the quantum vortex on a pole of magnet

(c) polar vortex rings of a Halbach array clearly shown by the ferrolens, middle column is the domain wall region.

Fig. 1. Quantum vortex flux of the field of various magnets as shown by the quantum magneto optic device ferrolens in real time. A RGB light ring was used on the perimeter of the lens showing the wire-frame individual quantum flux lines, of the field. (a) side field view of cube magnet under the ferrolens. North and South Poles are indicated and the domain wall of the magnet shown by the arrow. The two polar back to back vortices are clearly shown joint at the domain wall (b) polar field view inside the quantum vortex of a pole of cube magnet. Enormous holographic depth of the field shown by the ferrolens. A white light LED ring was used during the experiment since it produces the highest sensitivity for the ferrolens. The pole of the magnet is under the ferrolens facing it. The individual quantum flux lines shown are not criss-crossing but overlapping in 3D Euclidian space holographically shown by the lens. The ferrolens is magnetically transparent to both axial poles of the magnet therefore both poles of the cube magnet are projected simultaneously at its 2D surface [13] (c) A Halbach array under the ferrolens. Normally when magnet is under the ferrolens its quantum flux lines are extending outside and above the ferrolens surface before of course they all curl back towards the magnet's poles. However in this photograph taken, the two poles of the center magnet in the array shown are magnetically confined by the quantum polar vortices of the other magnets in the array left and right. Thus, vortex ring, torus, are formed on the two poles of the center magnet. The middle column at the center is the domain wall region of the dipole magnet.

*From the sample experimental* material taken of fig.1. above, the quantum vortex shaped magnetic flux existing in macro magnetic dipoles and Quantum Magnet is demonstrated and revealed for the first time by a quantum magneto optic device. Specifically, in fig.1(c) the North and South polar quantum flux forming perfect vortex rings or torus, is representing best this new Quantum Field of Magnets (QFM) geometry shown by the experiments and in compliance with the Maxwell equations [14] which demand zero divergence $\nabla \cdot B = 0$ **(1)** and full curl for magnetic fields. It is also remarkable how much evident with the ferrolens is, in fig. 1(a)(c), the domain wall or else referred as Bloch domain wall [15][16] region of the magnets. A region no more than 100 nm wide quantum effect, with the domain wall itself being a few atoms thick which is magnetically and spatially magnified by the lens and shown. Classically, in the domain wall region of a magnetic dipole field, the transition in polarity N-S or vice versa occurs. In the QFM vortex flux of a dipole magnet the domain wall is the joint between its two quantum flux N-S polar vortices which are axially connected and have a counter geometrical spin as shown in fig.1.

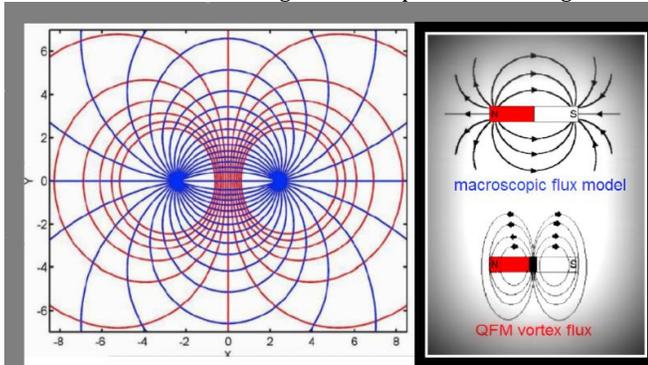

(a) The two existing flux modes of a magnetic dipole compared. The axial macroscopic E-field flux of the magnet and the spiral quantum vortex flux M-field of the magnet.

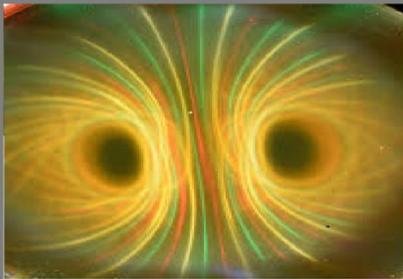

(b) The two polar quantum flux vortices (left and right black holes) of the QFM of a dipole magnet shown by the superparamagnetic 10 nm in size Fe3O4 magnetite nanoparticles in the ferrolens.

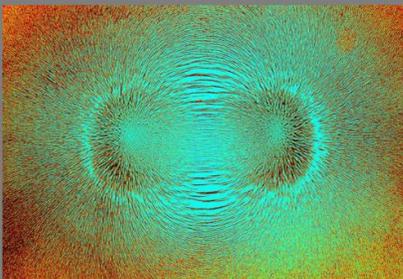

(c) The classic macroscopic field imprint of a dipole magnet as shown by macro field sensors like the mm sized Fe iron filings.

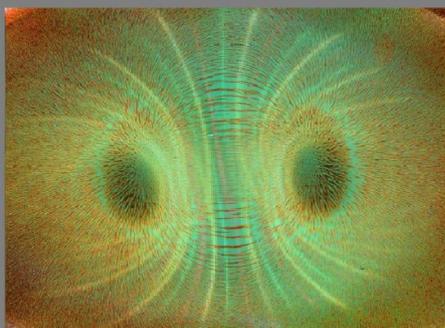

(d) Total QFM of dipole magnet (computer overlaid fields EM, b&c figures). Quantum vortex M-field and macroscopic iron filings E-field. The two polar vortices M-field (thicker lines) and the axial E-field (thinner lines). Imagine as two water vortices draining water displayed.

Fig. 2. Physical mechanism for explaining classical macroscopic E-field of macro magnetic dipoles with quantum vortex M-field (QFM) causality effect observed with the ferrolens. Fig.2(c) [17].

*This discovered,* QFM or else Quantum Magnet, vortex flux could also provide a possible physical mechanism for explaining the macroscopic field imprint of magnetic dipoles as shown in fig.2. The effect is similar as described by vortex hydrodynamics [18–20], quantum Bose-Einstein condensate ferrofluids [21] and general vortex model theory [22,23] very often encountered in nature from the quantum scale to the macro world [13]. The quantum vortex flux we call, M-field shown in fig.2(a) with red and fig. 2(b) as displayed in real time by the ferrolens, is possible responsible and the cause for the macroscopic axial flux we call E-field, fig.2(a) with blue and fig.2(c), we usually observe with macro field imaging sensors which have usually a mm scale size and are therefore susceptible to quantum decoherence phenomena. On the other hand ferrolens has very little quantum decoherence since it is using a nano scale imaging sensor (i.e. 10nm in average Fe3O4 magnetite nanoparticles).

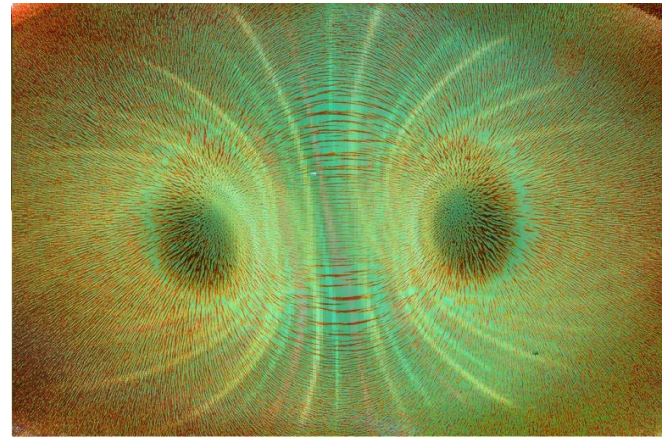

Fig. 3. Quantum Decoherence (QDE) effect mechanism responsible for the transmutation of the dipole net quantum field present on magnets *(see thick colored lines shown in the ferrolens )* into the classical macroscopic field imprint *(see tension field formed, brown iron filings thin lines, between the two black holes N-S poles of the magnet)*.

The total QFM of a dipole magnet with the combined E-M fields is displayed in fig. 2(d) and fig.3 using overlaid transparent images from the experiments to best demonstrate this causality effect of the two quantum polar vortices we call M-field as illustrated in fig.2a, acting like drains generating therefore the familiar, axial N-S classical macroscopic field we call E-field as illustrated in fig.2a, between the two poles *(i.e. see the two black holes in fig.2d and fig.3)*.

In fig.4 we see a sample of the single light source experiments we contacted with the ferrolens. A single LED was used as artificial lighting to activate the lens and was placed 1cm under the ferrolens at its center. The magnets used in the experiments were placed on top of the ferrolens. All observations and results obtained are in real time without any kind of processing. In this experimental configuration with a single light, a cloud pattern of the QFM-Magneton flux is displayed instead of the wire-frame flux pattern we get when using a LED ring. This allows us to observe more clearly and in a concise way the outer shell and general 360° outline geometry of the QFM flux in magnetic dipoles which is actually representing the field of a single magneton thus the Quantum Magnet. All experimental results obtained and a small sample is shown above in fig. 4a-d, demonstrate the same QFM geometry, independent of shape or material of magnet (i.e neodymium or ferrite) used in the experiments. Thus, a hemispherical field as subsequence of the quantum vortex field extending and curling in 3D Euclidian space. The two polar fields are joint axially at the domain wall forming the final spherical nature of magnetic fields. However as shown the two distinct N-S polar quantum vortices are also kept marginally separated by the domain wall.

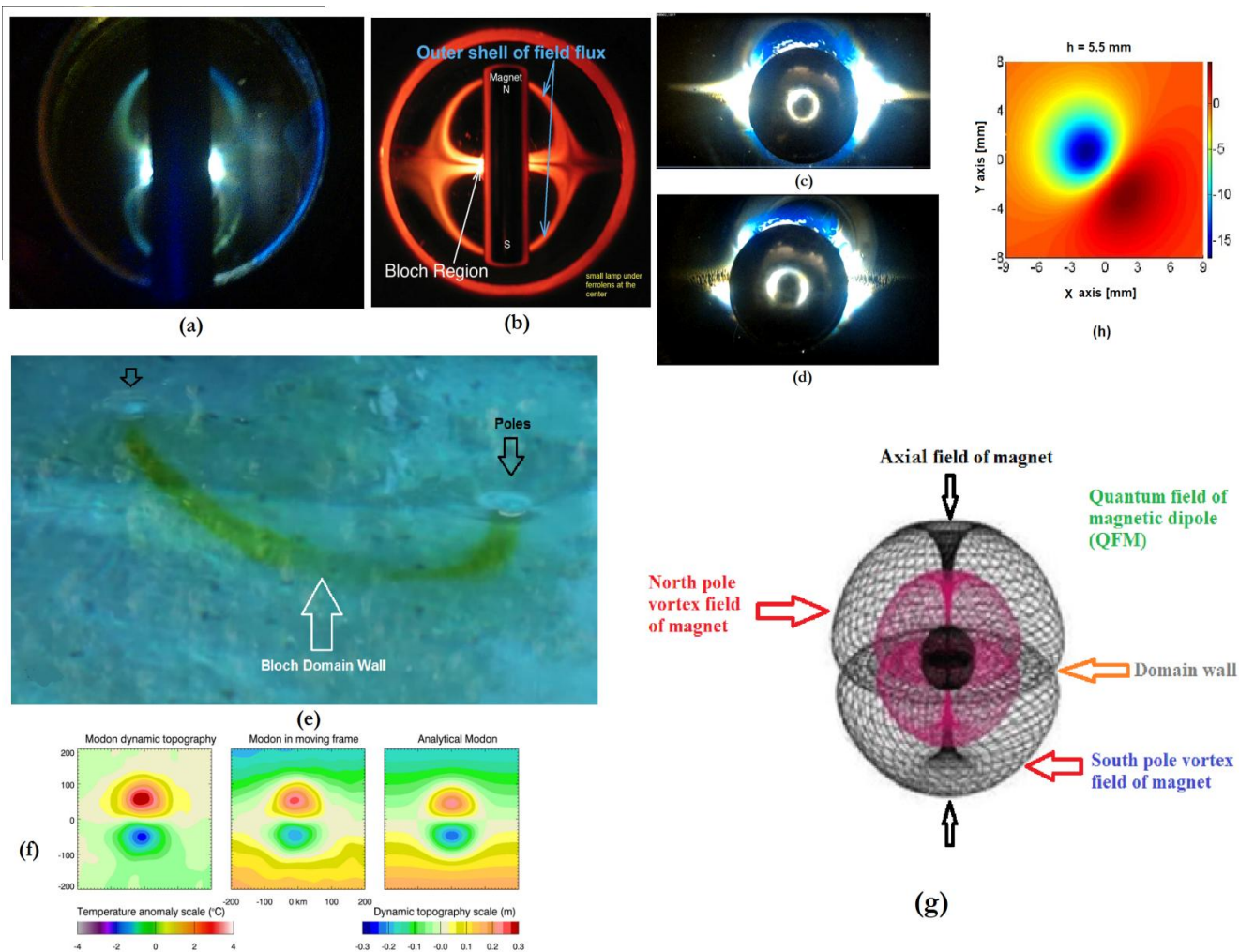

Fig. 4. Single light source ferrolens experiments and correlation of the total QFM-Magneton field of magnetic dipoles with classical vortex theory and Modons. (a) Cylindrical magnet placed on its side (side-field view) above the ferrolens with single white LED one cm under the lens at the center. Its two hemispherical polar fields outline is shown, magneton field. (b) Same experiment but with a yellow light LED. Look at inner theta θ pattern, magneton field. The outer perimeter is the rim of the lens. The hemispherical polar quantum field outline geometry is clearly again demonstrated with the domain wall joining at the middle and at the same time separating the two fields. (c) Same pattern seen on spherical magnet with the 2D Bloch domain wall disk at the middle. Upper part of the photograph is the North Pole and at the lower the South Pole. (d) Closer look at the domain wall of the spherical magnet. We can clearly see the separation of the polar quantum fields. (e) A modon. Counter spinning whirlpools [13] in a pool joining their vortices underwater as shown *(i.e. food coloring was used to make the effect visible)*. These vortices are joint together but at the same time keep their distance. (f) Velocity (flux) graphs of the modon [24]. Striking resemblance with the QFM of dipole magnets shown previously in (a)(b)(c) & (d). (g) Nested counter-torus hemispherical geometry of the total Quantum Field of Magnets (QFM) illustration. All of its geometrical field features are indicated including both of its flux field modalities, polar vortex flux and axial flux. The two polar quantum fields are joint at the domain wall in the middle as indicated. (h) Dipolar field of ferromagnet experimentally shown by another quantum sensor, scanning SQUID microscopy [25].

Surprisingly, we can observe the same field geometry and dipole behavior, in water modons [24,26] of two unaxially counter spinning whirlpools [13] as shown in fig. 4(e)(f). These whirls are joint and hold together underwater by the vortex made visible using food coloring resembling thus the domain wall of macro magnetic dipoles. Also, in the same time the whirlpool pair keeps its distance from each other avoiding merging. The same exact behavior which evidently is shown in all the sample QFM experiments, fig.1-4. *Similar dipolar field pattern* we can see also independently reported, from ferromagnets scanning SQUID magnetic microscopy, a quantum sensor, fig 4(h) [25].

Notice the striking resemblance of the analytic modon velocity (flux) graph in fig. 4f [24]. In fig. 4(g) we illustrate conclusively the observed total QFM geometry of magnetic dipoles as shown by the ferrolens and other experiments. A nested double counter torus joint hemispheres field geometry constituting a sphere. This field geometry is repeating in shells like an onion and extending fractally from center outwards in 3D Euclidian space. The two counter geometry North and South poles quantum vortex fields are indicated as well as the axial macroscopic field making up the final perpendicular EM fields of a dipole magnet with its domain wall essentially a 2D disk, in the middle of the sphere as illustrated in fig4(g).

A special macroscopic ferrolens at fig. 5(a), was constructed to demonstrate our point about quantum decoherence effect being responsible for reverting the quantum vortex field of a magnet QFM shown previously in fig. 1 to fig. 4, to its familiar axial macroscopic E-field imprint at fig. 5(b). The Fe iron field sensor particles we used are 40 μm in size, thus, x4,000 larger than the ones used in a normal quantum ferrolens, 10nm in size. The Fe iron particles are suspended in mineral oil carrier fluid and have a volume concentration percentage compared to the carrier fluid of about 30%. In a normal ferrolens this same percentage is no more than 0.75%. In fig. 5(b) a cylindrical Nd magnet is placed above the macro ferrolens and its N-S magnetic field is displayed.

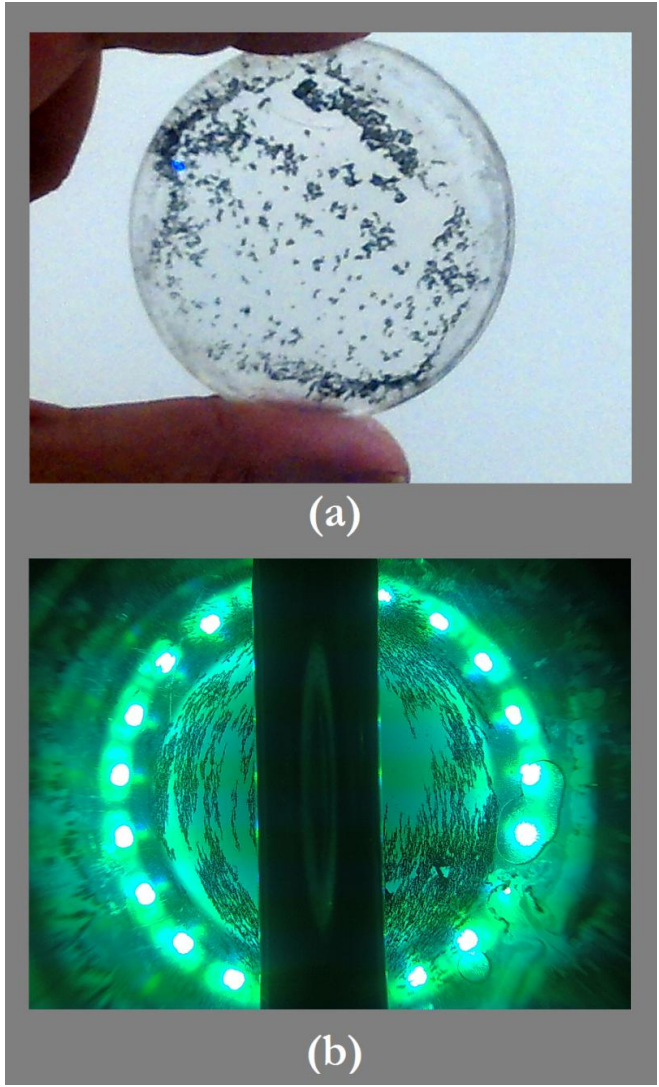

Fig. 5. Quantum decoherence effect demonstrated. (a) A 40μm Fe iron particles macroscopic ferrolens was made. The particles are suspended inside the lens in a mineral oil thin film. The particle size is here 4,000 times larger than the 10nm particle size used in a normal ferrolens. (b) Macroscopic familiar field imprint of a cylindrical magnet shown by the above macro ferrolens due to quantum decoherence effect of the actual net quantum vortex field present on the magnet previously displayed in fig.1,2,3&4, [13].

In the above macroscopic ferrolens experiment [13] shown in fig. 5, quantum decoherence rate δ [9,10,27] of the 40 μm iron Fe macro sensor particles inside the colloidal carrier fluid, about 200 μm in thickness encapsulated film, can be actually estimated by calculating the Brownian relaxation time $\tau_B$ [28,29] of the magnetic particles which is dominant at this relative macroscale size:

$$\tau_B = \frac{3V_B \eta_o}{kT} \quad Brownian\ relaxation\ time \quad (2)$$

$$V_B = \frac{4}{3}\pi(R+d)^3 \quad Brownian\ particle\ size \quad (3)$$

In equation (2), $V_B$, $\eta_o$ and $k$ are the particle volume, the dynamic viscosity of the carrier fluid and the Boltzmann constant approximately at $1.38 \times 10^{-23}$ respectively, with $T$ being the temperature set at 300K, room temperature for the purpose of this calculation. The spherical particle volume size $V_B$ in equation (3) is calculated for a radius R=D/2 at 20μm with the particles surfactant thickness d, being $R \gg d$ and therefore can be dismissed. The other specified value used in the calculation is $\eta_o$=2.4 cP thus, $2.4 \times 10^{-3}$ Kg/m/s for the carrier fluid mineral oil dynamic viscosity.

The quantum decoherence rate is estimated therefore to:

$$\delta = \tau_B^{-1} \cong 17 \times 10^{-6}\ Hz \quad (4)$$

Indicating that our *open system* ferrolens, is interacting strongly with the environment and behaving macroscopic, slowing down to a nearly hold, quantum fluctuations. Also since $\tau_B$ relaxation time is larger than experiment observation times, the ferrolens is now ferromagnetic as theory predicts [29].

In contrast, a superparamagnetic ferrolens with for example 10nm in diameter size magnetite $Fe_3O_4$ nanoparticles, would result due to its dominant Néel relaxation time $\tau_N$ over Brownian motion, at a $10^{-9}$ s time [28] and therefore to a $\delta \cong 10^9$ Hz, suggesting rapid magnetization fluctuations and that it is quantum active.

### 3.1 The irrotational net Quantum Field of Magnets (QFM) as actual cause for the cases of conservation in classical static macroscopic magnetic fields

In the ideal case of an undisturbed isolated by any external influence say for example a permanent magnet left alone and having a time invariant static field for a simply connected domain *(i.e. without any discontinuation in the field)* where there is no free electrons current present or any other external charge introduced, has a zero curl $\nabla \times \mathbf{H} = 0$ (5), and can be described by the negative gradient of its scalar potential, $\mathbf{H} = -\nabla \psi$ (6) and we also know that there is no work done and energy consumed by the magnet since there is no external charge introduced in field and in general there is no energy lost and the field is dormant. Although the above case does not classify as a vector force field, in the context of quantum mechanics since the magnet has aligned orbiting unbounded electrons interacting in its matter we can extent the definition of conservative fields for this case and say that, all three equivalent conditions described above are meet and the field is therefore described as conservative [30].

At the moment we introduce an external charge in the field of the magnet (i.e. real world condition since our magnet is an open system and interacting with the environment), it becomes now a vector force field and in general except special cases, the magnetic field of our permanent magnet is non-conservative in the presence of currents or time-varying electric fields by definition of the three criteria we described before and their equivalence proof meaning that when any one criterion holds there other two also hold and the field is described as conservative.

A conservative field should have a closed line integral (or curl) of zero [31]. Maxwell's fourth equation (Ampere's law) can be written here as:

$$\nabla \times \mathbf{B} = \mu_0 \mathbf{J} + \mu_0 \epsilon_0 \frac{\partial \mathbf{E}}{\partial t} \quad (7)$$

so we can see this will equal zero only in certain cases.

Magnetic force is also only conservative in special cases. The force due to an electromagnetic field is written

$$\mathbf{F} = q\mathbf{E} + q\mathbf{v} \times \mathbf{B} \quad (8)$$

For this to be conservative then $\nabla \times \mathbf{F} = 0$ and

$$\nabla \times \mathbf{F} = q\nabla \times \mathbf{E} + q\nabla \times (\mathbf{v} \times \mathbf{B}). \quad (9)$$

But from Faraday's law we know that
$\nabla \times \mathbf{E} = -\frac{\partial \mathbf{B}}{\partial t}$, so,

$$\nabla \times \mathbf{F} = -q\frac{\partial \mathbf{B}}{\partial t} + q\mathbf{v}(\nabla \cdot \mathbf{B}) - q\mathbf{B}(\nabla \cdot \mathbf{v}) + (\mathbf{B} \cdot \nabla)q\mathbf{v} - (\mathbf{v} \cdot \nabla)q\mathbf{B}$$

From Gauss's law for magnetism $\nabla \cdot \mathbf{B} = 0$ always, and for a single charge introduced $\nabla \cdot \mathbf{v} = \partial/\partial t(\nabla \cdot \mathbf{r}) = 0$.

Furthermore,
$(\mathbf{B} \cdot \nabla)\mathbf{v} = (\mathbf{B} \cdot \frac{\partial}{\partial t}\nabla)\mathbf{r} = 0$, so,

$$\nabla \times \mathbf{F} = -q\left[\frac{\partial \mathbf{B}}{\partial t} + \frac{\partial \mathbf{B}}{\partial x}\frac{\partial x}{\partial t} + \frac{\partial \mathbf{B}}{\partial y}\frac{\partial y}{\partial t} + \frac{\partial \mathbf{B}}{\partial z}\frac{\partial z}{\partial t}\right]$$

$$\nabla \times \mathbf{F} = -q\frac{d\mathbf{B}}{dt} \quad (10)$$

and the force is only conservative in the case of stationary static magnetic fields which also includes the case of our permanent magnet in our example used for this analysis.

Also we know that in the case of the single external charge introduced in the field of a permanent magnet, zero net work (W) is done by the force when moving a particle through a trajectory that starts and ends in the same point closed loop or else the work done is path-indepedent (i.e. not necessarily at the same point in the z-axis in 3D space, solenoid case) meaning equal amount of potential energy is converted to kinetic energy and vice versa and there is no energy loss in the system.

$$W \equiv \oint_C \vec{F} \cdot d\vec{r} = 0 \quad (11)$$

By the equivalence proof we already even with one criterion meet as we shown before, we should characterize the magnet interacting with the single charge as a conservative vector force field however in this case surprisingly the third criterion does not hold (12) and the equivalence proof is therefore broken

$$\vec{F} \neq -\nabla \Phi \quad (12)$$

since the force cannot be described as the negative gradient of any potential Φ we know. This anomaly directly implies that the single charge interacting magnetic field is not a conservative field but as a matter of fact not even a vector force field which is a contradiction of all we know and established about Electromagnetism.

A different approach and analysis is needed to establish that the static magnetic field of our magnet in our example, interacting with a charge is actual a conservative field thus path-independent, assuming that there is no work done as all experiments in the literature show and the equivalence only apparently does not hold for the third criterion as mentioned above in equation (12) simply because of our not complete 100% yet knowledge about Electromagnetism and not because the vector force field is not really conservative.

Actually this is not the only case in physics where a path-dependent thus non-conservative field can have zero curl. So we must be very carefully with the equivalence proof of the three criteria across different force fields and zero curl does not necessarily imply conservative although the opposite is always true thus, non zero curl fields cannot be conservative. Most velocity-dependent forces, such as friction, do not satisfy any of the three conditions, and therefore are non-conservative [32].

However the case we described of the static magnetic field interacting with a single charge, it is the only case in our knowledge that fulfills both of the zero curl and zero net work done criteria and fails in the third $\vec{F} = -\nabla \Phi$ (13) which is the most important and , in the context of the gradient theorem [33], it is an exclusive criterion and condition, thus, a vector field F is conservative if and only if it has a potential function $f$ with F=∇$f$ in general. Therefore, if you are given a potential function $f$ or if you can find one, and that potential function is defined everywhere, then there is nothing more to prove. You know that F is a conservative vector field, and you don't need to worry about the other tests we mention here. Likewise, if you can demonstrate that it is impossible to find a function $f$ that satisfies F=∇$f$, then you can similarly conclude that F is non-conservative, or path-dependent. As we stated before we do not yet know such a function potential to prove conservation and therefore another research lead and approach must be investigated.

So far our analysis using Maxwell theory for electromagnetism resulted that we cannot conclude that a static magnetic force vector field interacting with a single charge is conservative but without however rejecting the possibility.

Another large obstacle we have in characterizing the force field in our case and example as conservative by investigating alternative theories and analysis in the literature [34–37], is the conclusion coming directly from the gradient theorem and vector calculus in general saying that all conservative vector fields thus path-independent meaning that in our case, that the work done in moving a particle charge inside our magnetic field between two points is independent of the path taken and equivalently, if a particle travels in a closed loop the net work done by a conservative force field is zero, are also irrotational in 3D space and therefore must have a vanishing curl *(e.g. a vortex)* within a simply connected domain.

This last we can use as an exclusive criterion for conservative vector fields and all macroscopic magnetic fields according to Maxwell and experiments have nothing even near a vanishing curl even more are not irrotational by any means. Therefore, using this criterion no classical macroscopic magnetic field can ever be characterized as conservative.

However, our research presented herein shows an underlying hidden to the macro world net quantum dipole vortex field existing in every macro magnet as the actual causality field for generating the classical macroscopic magnetic field as a tensor function of the two polar quantum magnetic N-S polar vortices (fig. 3). This quantum magnetic field of magnets (QFM) is not present and visible at the macroscopic level due to Quantum Decoherence (QDE) phenomenon and mechanism.

This is a crucial finding that shows the actual vortex nature of magnetism in origin and suggesting that static magnetic fields at the quantum level are irrotational fields and therefore conclusively proving to be conservative.

Also notice, that unlike 2D space where a hole in the domain area of the field is not a simply connected domain which is a sub condition for the irrotational field being conservative as shown above, in 3D space a domain can have a hole in the origin center and still be a simply connected domain as long this hole does not pass all the way through the domain. This is also shown in all our experiments an observations made with the ferrolens of the field of permanent dipole magnets. The two N-S polar quantum vortex holes are not connecting and therefore the shown dipole vortex field is irrotational in a simply connected 3D domain of the total field of the magnet fig.4g and therefore conservative.

### 3.2 The logarithmic spiral arms of the QFM

From vortex theory [18] we know that a free vortex thus an irrotational vortex *(i.e. most common case of vortices occurring in nature)*, has a vanishing curl therefore a zero curl $\nabla \times \mathbf{B} = 0$ and all of its individual particle trajectories are *logarithmic spirals*. Also since it has a vanishing curl its trajectory progressively as the distance $r \to 0$ from the vortex origin (i.e. eye of vortex) diminishes, becomes a cirlce therefore a non-zero curl or else called the pole of the vortex.

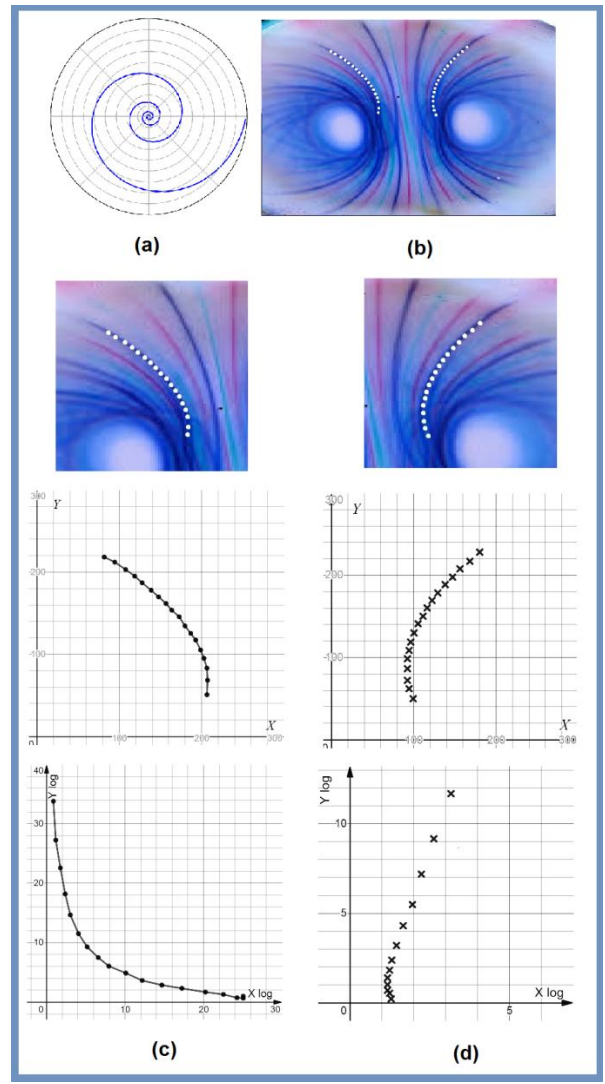

Fig. 6. (a) A logarithmic spiral (b) The QFM dipole vortex field of a magnet shown by our experiments using the Ferrolens same as fig.2b & fig. 3 but this time the negative of the photograph was used to select two spiral arm segments (see white dotted lines) of the QFM flux for numerical analysis. The xy Cartesian coordinates of these spiral segments were extracted using xy image digitizing technique. (c) A Left arm spiral segment of the QFM, used for numerical analysis. Up is its xy linear scale plot, down is its xy logarithmic scale plot (d) Right arm spiral segment and its corresponding data plots.

The experimental data of fig. 3 dipole vortex photograph was used in the numerical analysis by extracting the xy Cartesian coordinates of two QFM flux lines segments using image digitization technique and the corresponding xy plots of these were drawn to decide if these spiral arms are logarithmic.

In a logarithmic spiral its polar equation is,

$$r = ae^{b\theta} \tag{13}$$

where $r$ is the distance from the origin {0,0} of the vortex, $\theta$ is the angle from the x-axis, and $a$ and $b$ are arbitrary constants.

Also its *xy Cartesian coordinates are equal to,*

$$x = r\cos\theta = a\cos\theta e^{b\theta} \quad (14)$$

$$y = r\sin\theta = a\sin\theta e^{b\theta} \quad (15)$$

and the change of its radius is,

$$\frac{dr}{d\theta} = abe^{b\theta} = br \quad (16)$$

In fig.6c and fig.6d we can see the plots derived from the numerical analysis of the two spiral arms of the QFM vortex field shown by the Ferrolens. The first plots of each column are drawn with linear scale xy axes (*see links for the extracted numerical values* https://tinyurl.com/y3nekkub and https://tinyurl.com/yyojfz57), the second plots below on each column are the same plot numerical data but now the xy axes used have logarithmic scale (https://tinyurl.com/yyln6jsn and https://tinyurl.com/y554k3nt ). We see that in the second case where the logarithmic scale axes are used, the plots give exponential functions and therefore the original spiral arms of the QFM of the magnet shown by the Ferrolens are logarithmic spirals allowing also a minute optical distortion depending on the view angle from the lens *(see log-log verification plots at* https://tinyurl.com/yxh8r9cg , https://tinyurl.com/y5eq7mo3 *accordingly).*

### 3.3 Quantum Magnet

A novel observation has been made that the collective net Quantum Magnetic Field of the Magnet-Magneton (QFM) consists of a dipole vortex shaped magnetic flux geometrical pattern responsible for creating the classical macroscopic N-S field of magnetism as a tension field between the two polar quantum flux vortices North and South poles. A physical interpretation of this quantum decoherence mechanism observed was analyzed and presented, showing physical evidence of the quantum origin irrotational and therefore conservative property of magnetism and also demonstrating that ultimately magnetism at the quantum level is an energy dipole vortex phenomenon *(i.e. we known that magnetic flux lines are made up of virtual photons flow).*

In this point it is important to explain that the authors here by the data and the analysis presented, never reported or implied that the magnetic flux geometrical vortices of permanent magnets shown on the ferrolens are a result of topological defects, giant skyrmions, thus induced by the superparamagnetic thin film hermitically sealed inside the lens. But are quite accurately representations of the actual **net Quantum Magnet** field (QFM) we discovered present in all macroscale permanent magnets [1]. To our knowledge, there were never reported by the literature ever existing cm in size, stable magnetic skyrmions [39] of any type superfluidic or superconductive let alone at room temperatures. Therefore the quantum magnetic field imprint on the ferrolens is exclusively induced by the external macro magnet under observation. Additionally, the 25-50 μm thick depending the ferrolens construction, thin film colloidal magnetite $Fe_3O_4$ ferrofluid-kerosene mixture used inside the ferrolens and sealed to prevent evaporation, is non-magnetohydrodynamic, meaning it is an electrical insulator and the nanoparticles have antistatic coating [1] [2].

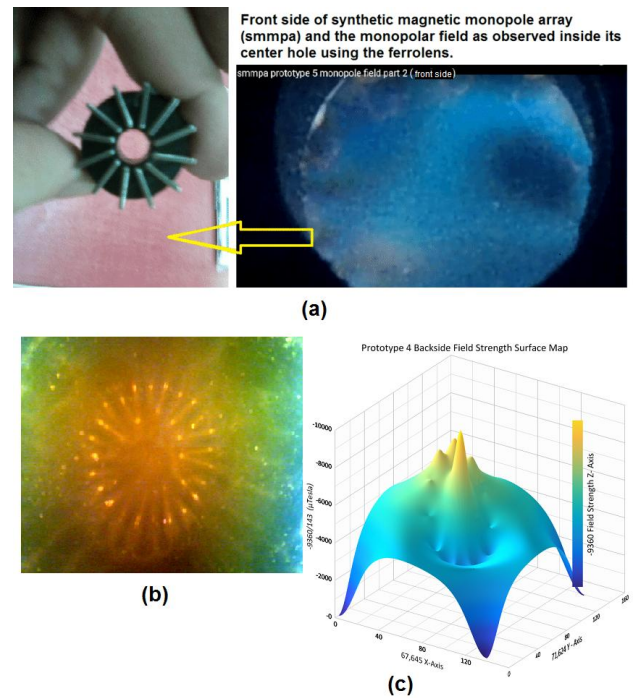

Fig. 7 (a) Synthetic Magnetic Ring Array and its quantized vortex pole at the center shown on a ferrolens (b) The 12 perimeter magnetic plates used in the array, quantized flux shown by the ferrolens (c) Surface map B-field strength of the ring array measured with a 3-axis magnetometer.

In order to demonstrate herein that the ferrolens is driven exclusively by the external magnetic field, in fig. 7(a), a special magnetic ring array [1] was constructed, a magnetic flux twister, consisting of twelve 10mm X 1mm square magnet plates inserted vertically in a 3D printed PVC frame. The combined quantized magnetic flux of the twelve magnets in the ring array, produces a circular type magnetic moment clockwise or anticlockwise depending the magnets placement but because the magnets have a skewed angle to each other inside the ring it results to an elliptical vortexing magnetic moment with a vanishing curl and inducing to the ferrolens nanoparticles a progressively incremental angular velocity which its magnetic imprint is accurately picked up by the ferrolens as shown *(i.e. see the field at the center of the ring array which is placed on top of the ferrolens)*. In fig. 7(b) we see the quantized flux of the twelve magnets ring placed under the ferrolens and because their very small 1mm thickness and skew angles, the magnetic flux is forced to curl and wave around and in between the magnets in the ring almost like a sinusoidal wave function. In fig. 7(c) we see the magnetic field B intensity surface map of the synthetic magnetic ring array as measured with 3-axis magnetometer, placed 1cm away from the face of the ring array (measured values found in the link https://tinyurl.com/yypmhglv). Each hump on the perimeter is a magnetic plate of the ring with the net field single Pole of the ring formed at the center. The elliptical perimeter and downward slope of the surface map indicating the vortexing external magnetic field generated by the synthetic magnetic ring array prototype.

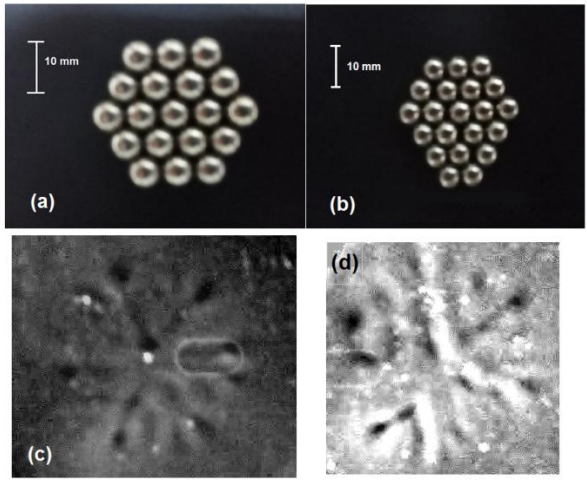

Fig. 8 (a) Nineteen, 5mm sphere magnets triangular lattice (b) Twenty-one 5mm sphere magnets triangular lattice (c) & (d) Their corresponding quantized fields and magnetic flux vortices as shown by the ferrolens. The triangular phases formed, of the quantized magnetic field lattices are also shown. Some dust and reflection photographic artifacts are present.

So far, we have demonstrated and analyzed the dipole **Quantum Magnet Field (QFM)** of macro permanent magnets shown by the ferrolens, a quadrupole face part from a Halbach magnetic array and the quantized field of a twelve magnets synthetic ring array, magnetic flux twister.

In fig. 8 experiment, we placed different triangular lattices configurations of 5mm in diameter small sphere dipole permanent magnets under the ferrolens with a black cupboard paper inserted in between for increased contrast of the magnetic field imprint. Notice, that the net quantized field of the lattices induced, is relative small in strength, near the sensitivity limit of the used ferrolens in the experiment. This requires at least minimum field strength of 15 mT to start to display. Therefore the photographs taken of the field shown by the ferrolens are very fade and were software enhanced for increased contrast. The displayed, *see fig. 8(c)&(d),* in real time by the ferrolens, artificially induced by the magnetic lattices, macroscopic-near microscopic quantized vortices and their triangular field phases interactions, demonstrates the capability of the lens to pick up and display macroscopic externally induced quantized magnetic flux and vortices. Similar to the generated quantized vortex lattices observed in BECs [21] [39]. However, we have to stress out here that, this quantum magnetic optic device must not in any way assumed of being capable and responsible, for the generation of these stable quantized vortices due topological defects in the medium since it does not exhibit any superfluidic nor superconductive properties at room temperatures as explained before. In other words, the Ferrolens device can show macroscopic quantized magnetism and vortices but cannot generate them. The ferrolens optic display device is an electric insulator at all temperatures therefore not a superconductor. In addition we tested rigorously for superfluidic behavior concerning quantization of magnetic vortices generation at different values of induced angular momentum [21] [39] without any observable effect.

The device remains neutral and optical isotropic unless disturbed by an external magnetic field *(see link https://tinyurl.com/y26yuru5)* and returns to its previous isotropic idle condition instantly after the magnet is removed. The encapsulated thin film of ferrofluid inside the ferrolens in this state, does not flow, but exists in a balanced state of equilibrium no matter what position the cell is oriented. The nanoparticles inside the ferrolens do not settle with gravity. Moreover, when a magnet is fastened at the ferrolens surface the magnetic field displayed is unaffected by any motion of the ferrolens device.

Therefore, from all the experimental results and analysis presented herein and our previous research [1][2], we come to a conclusion and deduction by elimination process, that the ferrolens is probing and accurately displaying the actual net collective **Quantum Magnet Field (QFM)** present in every macroscopic size magnet and otherwise masked at the macroscale and macroscopic field sensors due Quantum Decoherence (QDE) phenomenon and also recognize that the whole phenomenon warrants further investigation in the future.

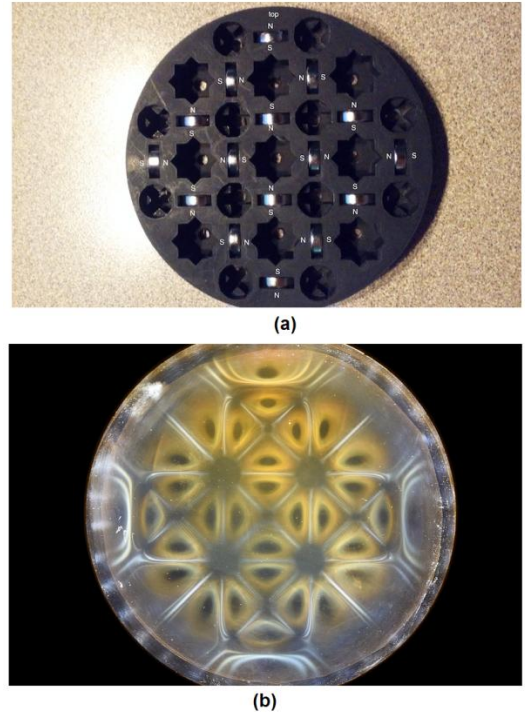

Fig. 9 (a) Forced orientation magnetic array (b) Field shown with ferrolens.

Finally, because the field images obtained in fig. 8 from the freely oriented 5mm sphere magnet lattices are at the limit of this ferrolens sensitivity, we used in this particular experiment and in order to demonstrate the actual field analytical display capability of the ferrolens device, we constructed a forced polarity orientation, magnetic array shown in fig. 9(a) using much stronger sixteen Neodymium N42 10mmX4mm disk magnets. The magnetic arrays was placed under the ferrolens in contact and a black cupboard paper was inserted in between to mask the view of the magnets and make sure that only the field is displayed and also increase the contrast.

The field real time display of this magnetic array by the ferrolens is shown in fig. 9(b) and the orientation of the individual magnetic moments N-S in the array is indicated in fig. 9(a). Each vortex, black hole, shown is a Pole of a magnet in the array and each dipole vortex (i.e. vortices pair) two joint hemispheres is a magnet. A close inspection of fig. 9b reveals even the individual separated magnetic flux lines geometrically vortexing around the Poles. We can also observe the domain walls triangular phases formed, see grid of white flux lines, due the interaction of the magnets inside the array. Once again the **Quantum Magnet Field (QFM)** emerges here for each single magnet in the array as we demonstrated previously in fig. 4a&b for the dipole magnet, thus the two hemispherical polar fields or polar vortices, axially joint at the domain wall in the middle.

The four gray spots in the center forming a square are field cancelation areas in the magnetic array formed by antiparallel pairing of the fields of the magnets in these areas.

Last but not least, we observe the four corner Poles in the magnetic array to be opened up vortices as expected since these particular Poles are very little to none interacting with the rest of the magnets in the array and therefore are not magnetically confined.

## 4. CONCLUSIONS

We have herein introduced, presented and analyzed by experimental evidence and theoretical analysis, the existence of a second hidden quantum vortex flux residing in every macro magnetic dipole *(i.e. we experimented only with macro magnets and not with any microscopic or quantum sized magnetic dipoles)* besides the macroscopic axial flux observed in magnetic dipoles. This second quantum vortex field is masked from the macrocosm and macroscopic sensors due to quantum decoherence (QDE) and is showing up only when a quantum magnetic field imaging device is used with nano sized sensors. In a way we can say that the ferrolens is the quantum version of the classical Faraday's iron filings experiment.

Therefore we made the novel observation that the collective net QFM-Magneton field consists of two quantum magnetic flux vortices geometrical patterns, joint back to back, and with each vortex residing on each pole of the Quantum Magnet as shown in fig.2a & fig.3 and in their outline form in fig.4(a)(b), present in every macro magnetic dipole besides its macroscopic classical N-S axial flux field.

A valid correlation was made between our experimental results and analysis with SQUID magnetic microscopy, quantum Bose-Einstein condensate ferrofluids, general vortex theory and specifically hydrodynamics modon dipole phenomenon occurring in nature, with striking similar effects with our observations with the ferrolens of magnetic dipoles and also latest quantum dots research for the magnetic model of the electron [38].

This *newly* experimentally *observed* vortex structure QFM in dipole magnets and Quantum Magnet found is complying to the Maxwell equation (1) (*i.e. $\nabla \cdot B = 0$* ) for zero divergence and exhibits full curl when observed with the ferrolens. Nevertheless, it could also potentially provide a mechanism as shown, to explain the macroscopic classical magnetic field imprint of magnets as a subsequent quantum decoherence effect and prove therefore the actual we believe, ultimately vortex nature of magnetism we have researched by analyzing the individual magnetic flux lines geometry of the magneton observed with the ferrolens and concluding that magnetism is a dipole energy vortex phenomenon (i.e. magnetic flux lines are made up of virtual photons flow).

In fig. 5 shown experiments, we demonstrated clearly that quantum decoherence is taking place here and is responsible for reverting the reported by us masked, quantum vortex field QFM of macro dipole magnets to the familiar macroscopic N-S field. Also, theoretical calculations were made to confirm that quantum decoherence is the cause for this phenomenon.

Also, a numerical analysis of the experimental data was carried out to confirm that the QFM vortices shown by the Ferrolens are natural logarithmic spirals thus free irrotational vortices.

Additionally, a theoretical analysis was presented showing that in special cases where a magnetic force vector field is to be considered as conservative it must be necessarily irrotational with a vanishing curl therefore confirming our findings about the actual quantum origin, elementary dipole vortex shaped field and nature of magnetism.

Besides the experiments carried out with normal dipole macroscale magnets, in section 3.3 we demonstrated and analyzed also experimental results obtained from magnetic arrays and lattices and concluded that these results, vortices shown on the Poles of magnets, are not due topological defects of the ferrolens medium, quantum field imaging sensor or other phenomena we previously investigated [1]. Thus, superfluid or superconductive magnetic quantized vortices shown usually in Bose-Einstein Condensates, but exclusively due to the **Quantum Magnet** collective net aligned magnetic moments and collective field *(see fig. 4a&b)* of the unpaired electrons inside the magnet material projected into the Ferrolens quantum magnetic optic device and successfully displayed. The collective quantum field of the magnet resembles closely to the stationary intrinsic magnetic dipole field of the single electron called Quantum Magnet or else Magneton.

Therefore, the ferrolens acts like a quantum optic microscope and our method presents to us potentially a unique opportunity for the first time to map the magnetic flux geometry *(see fig. 1, fig. 2&3 and fig. 4g)* and dynamics of the single elementary Magneton-Quantum Magnet thus the stationary magnetic field of a single electron. In essence what we see with the Ferrolens of the field of a macroscopic permanent magnet is, we look directly at the field of a giant sized stationary magneton.

This macroscale projected quantum phenomenon reported first here we believe warrants further investigation in the near future with the potential of new discoveries and new physics.

**Acknowledgements.** The authors acknowledge their colleagues, Dr. Michalis Tatarakis, Dr. Manolis Maravelakis, and Dr. Stelios Kouridakis Technological Educational Institute of Crete Academic staff members, for their support. Also Mr. Michael Snyder MSc, Timm Vanderelli independent researchers USA and Dr. Alberto Tufaile University of São Paulo Brazil for their support.
**Disclosures.** All authors contributed equally to the work. The authors declare that there are no conflicts of interest related to this article.

**Appendix:** Supplementary material to this article can be found online at https://tinyurl.com/y3je49tb [13].

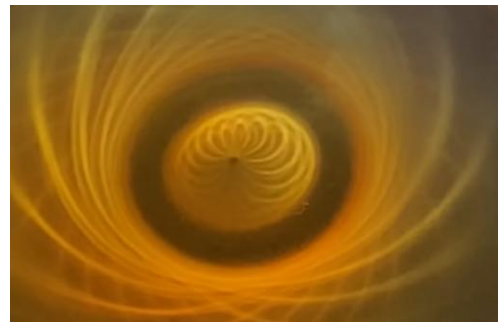

Graphical Abstract: Quantum magnetic vortex field (QFM) of a neodymium ring magnet shown by the ferrolens in real time. The field on the perimeter of the ring magnet extends and curls on the outside holographically shown by the ferrolens with the field inside the ring magnetically confined and forming a torus.